# Accretion discs around black holes: two dimensional, advection cooled flows

Igor V. Igumenshchev[1], Xingming Chen & Marek A. Abramowicz

*Department of Astronomy & Astrophysics, Göteborg University and Chalmers University of Technology, 412 96 Göteborg, Sweden*

**ABSTRACT**
Two-dimensional accretion flows near black holes have been investigated by time-dependent hydrodynamical calculations. We assume that the flow is axisymmetric and that radiative losses of internal energy are negligible, so that the disc is geometrically thick and hot. Accretion occurs due to the overflow of the effective potential barrier near the black hole, similar to the case of the Roche lobe overflowing star in a binary system. We make no pre-assumptions on the properties of the flow, instead our models evolve self-consistently from an initially non-accreting state. The viscosity is due to the the small-scale turbulence and it is described by the $\alpha$-viscosity prescription. We confirm earlier suggestions that viscous accretion flows are convectively unstable. We found that the instability produces transient eddies of various length-scales. The eddies contribute to the strength of the viscosity in the flow by redistributing the angular momentum. They also introduce low amplitude oscillatory variations which have a typical frequency about 100 $M_\odot/M$ Hz for a system of mass $M$. This may be relevant to the high frequency ($\sim 4 - 10$ Hz) quasi-periodic oscillations observed in the Galactic black hole candidate X-ray sources.

---

[1] Also Institute of Astronomy, 48 Pyatnitskaya Street, Moscow, 117810, Russia

# 1 INTRODUCTION: ADVECTION IN THIN, SLIM AND THICK DISCS

One of the astrophysical motivations for the present work is the recently recognized importance of the advective cooling in some accretion flows. Advection dominated accretion discs are hotter and less effective in converting the gravitational energy of accreted matter into radiation than the standard Shakura-Sunyaev discs, and thus have very sub-Eddington luminosities even for quite 'normal' accretion rates. For this reason Abramowicz & Lasota (1995) called them the *secret guzzlers*. Narayan and his collaborators pointed out recently that each object which is suspected to contain an accreting black hole and appears to be hot and underluminous could be a secret guzzler: advection dominated, hot accretion disc. This class of objects includes the black-hole X-ray transients like A0620-00 (Narayan, McClintock, & Yi 1995), the Galactic Center source Sgr A* (Narayan, Yi, & Mahadevan 1995), and the LINER NGC 4258 (Abramowicz et al. 1995b). For all three cases there are excellent spectral fits to that observed.

Accretion flows which are predominantly cooled by advection cannot be assumed to be geometrically very thin and so it is interesting to construct two dimensional models of them, without performing the usual vertical integration. On the other hand, radiative losses are not a dominant cooling mechanism for these flows and therefore they may be neglected in the first approximation, especially when one is exploring the global hydrodynamical behaviour and leaves the calculation of the emerging electromagnetic spectrum to a future work. For this reason, in this paper we study two-dimensional accretion flows with negligible radiative losses.

In the remaining part of this Introduction we follow the recent review by Abramowicz & Lasota (1995), quoting almost directly their short and concise description of the development of the theory of advection dominated flows. In the standard model of *thin accretion discs* around black holes (Pringle 1981) it is assumed that the ratio of the disc thickness $H = H(R)$ to the radius $R$ is very small, $H/R \ll 1$. The heat is generated at a rate $Q^+(R)$ by viscous stresses, and removed by radiation losses at a rate $Q^{\rm rad}$. In the steady state a *local* heat balance, $Q^+ = Q^{\rm rad}$, holds. The advective flow of heat, $Q^{\rm adv}$, caused by the accretion flow of matter, is always present. However, this heat flow is negligibly small for thin discs because $H/R \ll 1$ and one may show that,

$$Q^{\rm adv} \sim \left(\frac{H}{R}\right)^2 Q^+. \tag{1.1}$$

On the other hand, the advective cooling *must* become important in situations when $H/R$ is not small. It was noticed already a long time ago by Begelman (1978, 1979) and Begelman & Meier (1982) that in the case of a spherical (or a quasi-spherical) accretion a significant amount of heat may be trapped inside matter, advected into the central black hole and lost. Kozłowski, Jaroszyński, & Abramowicz (1978) and Abramowicz, Calvani, & Nobili (1980) calculated the rate of the advective leaking of heat from a *thick accretion disc* due to the relativistic Roche lobe overflow, and Abramowicz (1981) demonstrated that the leaking stabilizes both viscous and thermal modes at the disc's inner edge. Theory of optically thick accretion flows with a significant advection cooled was established by Abramowicz, Lasota, & Xu (1986) and Abramowicz et al. (1988) and it is now rather well-known as the *slim accretion disc* approach.

Although the corresponding theory of optically thin accretion discs with significant advection was established only very recently, it is rooted in an old difficulty, which was embarrassing theorists for almost two decades. Indeed, already two decades ago Shapiro, Lightman & Eardley (1976) constructed a two-temperature, hot ($T_{\rm e} \sim 10^9$ K), optically thin model of accretion disc to explain the hard X-ray spectrum of Cygnus X-1. The difficulty was that their model had $H/R \gtrsim 0.2$ contrary to the assumption $H/R \ll 1$ which was used in its construction. Worse than that, the model was thermally unstable. Numerous more elaborated versions of the model have been constructed later by several authors, but they always suffered from the same difficulty. It was recognized only recently that the *dominant* cooling mechanism in this kind of accretion flows is the global advection. When advection is included, the difficulty disappears. The first step in the understanding of this point was done by Narayan & Popham (1993) who discussed the importance of advection in optically thin accretion disc boundary layers. Later, Narayan & Yi (1994, 1995) constructed several self-similar models of optically thin accretion discs which were two-temperature and the bremsstrahlung and Comptonized synchrotron radiation was included. Similar optically thin, hot, advection dominated solutions with a non-relativistic bremsstrahlung were independently calculated by Abramowicz et al. (1995a), who have demonstrated that these new solutions are viscously and thermally stable. Chen et al. (1995) gave a unified description of black hole accretion flows and discussed the role of advective cooling from a general perspective.

In this paper we calculate a number of two-dimensional hydrodynamical models of disc accretion into black hole which are relevant to advection dominated accretion flows. We study the innermost regions of accretion disc, ignoring radiative losses of internal energy of the accreted matter. This may be a quite satisfactory approximation in the case of advection cooled flows. It is known from the theory of optically thick slim accretion discs that advection is a dominant cooling mechanism and the local radiative cooling is relatively unimportant when the mass accretion rate, $\dot{M} \gtrsim L_{Edd}/c^2$ (where $L_{Edd} = 4\pi cGMm_p/\sigma_T$ is the Eddington luminosity, $m_p$ is the proton mass, and $\sigma_T$ is the Thomson cross-section). For this reason, we expect that our models which ignore the radiative losses are physically correct in this regime. The basic equations of the problem are given in Section 2. The static torus which we use as the initial condition for our numerical models is shortly described in Section 3, and a brief description of the numerical method is given in Section 4. Numerical results are presented in Section 5, where the characteristic properties of the accretion flow are also discussed. Final discussion and projects for future investigations are contained in Section 6.

## 2 TWO DIMENSIONAL ACCRETION DISC MODELS

Realistic accretion flows around black holes call for general relativistic, three-dimensional, time-dependent numerical models with dissipative magnetohydrodynamics and radiative transfer. They should self-consistently describe turbulence at all relevant length-scales. It is not possible to construct such models yet, and thus various approximations are adopted by authors who study the subject. The present study considers axisymmetric two-dimensional flows in which the cooling is dominated by the global advection, and for this reason we mention here only these previous works which in our opinion are directly relevant to the present study. In particular, we do not mention works based on the vertical integration.

## 2.1 Some previous works on two-dimensional models of accretion flows

Theory of two dimensional accretion discs (tori) around black holes has been developed quite independently from the problem of advective cooling. Analytic models of static tori have been introduced about twenty years ago by Fishbone & Moncrief (1976) and Abramowicz, Jaroszyński, & Sikora (1978). They have been further studied by Kozłowski et al. (1978), Jaroszyński, Abramowicz, & Paczyński (1980), Paczyński & Wiita (1980), Abramowicz et al. (1980), and Rees et al. (1982). Two-dimensional numerical models of accretion discs have been studied only recently. For example, Kley & Lin (1992) studied accretion discs around white dwarfs, and Różyczka, Bodenheimer, & Bell (1994) constructed protoplanetary disc models around young solar-type stars. Hawley, Smarr, & Wilson (1984ab) have developed a general relativistic numerical code in the Kerr geometry, and simulated the formation of gas pressure supported tori around black holes. Eggum, Coroniti, & Katz (1987) have calculated radiative hydrodynamical models of sub-Eddington accretion discs and later they studied super-Eddington black hole accretion flows with jet formation (Eggum, Coroniti, & Katz 1988). Most recently, Papaloizou & Szuszkiewicz (1994) constructed two-dimensional stationary transonic accretion discs in order to compare them with vertically integrated one-dimensional models.

In several cases the 2D problem was reduced to an effectively 1D one by adopting some additional assumptions of physical or mathematical nature. Paczyński (1980) assumed that accretion goes in a narrow layer at the surface, while Paczyński & Abramowicz (1982) and Różyczka & Muchotrzeb (1982) assumed that all the accretion is restricted to the equatorial plane. Self-similar solutions for two-dimensional accretion have been examined, for example, by Begelman & Meier (1982), Liang (1988), and Narayan & Yi (1995).

## 2.2 Equations for two-dimensional accretion flows

We use cylindrical coordinates $(r, z, \phi)$. The accretion disc is assumed to be non self-gravitating and axisymmetric with the rotation axis coincident with the $z$-axis. The flow is adiabatic, in particular, no local radiative cooling is included. We describe the gravitational field of the black hole in terms of Newtonian hydrodynamics, with the gravitational potential introduced by Paczyński & Wiita (1980),

$$\Phi(r, z) = -\frac{GM}{R - R_\mathrm{G}}. \tag{2.1}$$

Here $R = (r^2 + z^2)^{1/2}$, $R_\mathrm{G} = 2GM/c^2$ is the gravitational radius of the black hole and $M$ its mass. The pseudo-Newtonian potential (2.1) for the black hole gravitational field was used by numerous researchers before. It was found that for static tori it gives qualitatively the same results as the correct general relativistic calculations and that the quantitative agreement is typically better than a few percents.

The equations describing the problem consist of:
the continuity equation,

$$\frac{\partial \rho}{\partial t} + \frac{1}{r}\frac{\partial}{\partial r}(r\rho v_r) + \frac{\partial}{\partial z}(\rho v_z) = 0, \tag{2.2}$$

the equations of motion,

$$\frac{\partial}{\partial t}(\rho v_r) + \frac{1}{r}\frac{\partial}{\partial r}(r\rho v_r^2) + \frac{\partial}{\partial z}(\rho v_r v_z) = -\frac{\partial P}{\partial r} - \rho\frac{\partial \Phi}{\partial r} + \rho\frac{\ell^2}{r^3} + q_r, \tag{2.3}$$

$$\frac{\partial}{\partial t}(\rho v_z) + \frac{1}{r}\frac{\partial}{\partial r}(r\rho v_z v_r) + \frac{\partial}{\partial z}(\rho v_z^2) = -\frac{\partial P}{\partial z} - \rho\frac{\partial \Phi}{\partial z} + q_z, \tag{2.4}$$

$$\frac{\partial}{\partial t}(\rho \ell) + \frac{1}{r}\frac{\partial}{\partial r}(r\rho \ell v_r) + \frac{\partial}{\partial z}(\rho \ell v_z) = rq_\phi, \tag{2.5}$$

the energy equation,

$$\frac{\partial}{\partial t}(\rho \varepsilon) + \frac{1}{r}\frac{\partial}{\partial r}(r\rho \varepsilon v_r) + \frac{\partial}{\partial z}(\rho \varepsilon v_z) = -P \operatorname{div} \vec{v} + Q, \tag{2.6}$$

and the equation of state,

$$P = (\gamma - 1)\rho \varepsilon. \tag{2.7}$$

Here $\vec{q} = (q_r, q_z, q_\phi)$ is the viscous force, and $Q$, $\ell$, and $\varepsilon$ are the heat dissipation function, the specific angular momentum, and the specific internal energy respectively. All other symbols have their standard meanings (Tassoul 1978). The viscous force and the dissipation function in cylindrical coordinates can be expressed as (see Tassoul 1978),

$$q_r = \frac{1}{r}\frac{\partial}{\partial r}(r\tau_{rr}) + \frac{\partial \tau_{rz}}{\partial z} - \frac{\tau_{\varphi\varphi}}{r}, \tag{2.8a}$$

$$q_z = \frac{1}{r}\frac{\partial}{\partial r}(r\tau_{zr}) + \frac{\partial \tau_{zz}}{\partial z}, \tag{2.8b}$$

$$q_\varphi = \frac{1}{r}\frac{\partial}{\partial r}(r\tau_{\varphi r}) + \frac{\partial \tau_{\varphi z}}{\partial z} + \frac{\tau_{r\varphi}}{r}, \tag{2.8c}$$

$$Q = 2\mu \left(D_{rr}^2 + D_{zz}^2 + D_{\varphi\varphi}^2 + 2D_{rz}^2 + 2D_{r\varphi}^2 + 2D_{z\varphi}^2\right) - \frac{2}{3}\mu (\operatorname{div} \vec{v})^2. \tag{2.8d}$$

Here

$$D_{rr} = \frac{\partial v_r}{\partial r}, \quad D_{zz} = \frac{\partial v_z}{\partial z}, \quad D_{\varphi\varphi} = \frac{v_r}{r},$$

$$D_{rz} = \frac{1}{2}\left(\frac{\partial v_z}{\partial r} + \frac{\partial v_r}{\partial z}\right), \quad D_{r\varphi} = \frac{r}{2}\frac{\partial}{\partial r}\left(\frac{\ell}{r^2}\right), \quad D_{z\varphi} = \frac{1}{2}\frac{\partial}{\partial z}\left(\frac{\ell}{r}\right)$$

are the components of the shear tensor, and

$$\tau_{rr} = 2\mu D_{rr} - \frac{2}{3}\mu \operatorname{div} \vec{v}, \quad \tau_{zz} = 2\mu D_{zz} - \frac{2}{3}\mu \operatorname{div} \vec{v}, \quad \tau_{\varphi\varphi} = 2\mu D_{\varphi\varphi} - \frac{2}{3}\mu \operatorname{div} \vec{v},$$

$$\tau_{rz} = \tau_{zr} = 2\mu D_{rz}, \quad \tau_{r\varphi} = \tau_{\varphi r} = 2\mu D_{r\varphi}, \quad \tau_{z\varphi} = \tau_{\varphi z} = 2\mu D_{z\varphi},$$

are the components of the viscous stress tensor. In addition,

$$\operatorname{div} \vec{v} = \frac{1}{r}\frac{\partial}{\partial r}(rv_r) + \frac{\partial v_z}{\partial z},$$

and $\mu$ is the dynamical viscosity coefficient. We model the small-scale turbulence by assuming that it introduces viscosity, which can be described by the $\alpha$ prescription (Shakura & Sunyaev 1973). In the case of two-dimensional flows (not vertically thin), the $\alpha$ prescription takes the form,

$$\mu = \alpha \rho \frac{c_s^2}{\Omega_k}, \quad (2.9)$$

where $\alpha$ is a constant, $c_s = (\gamma P/\rho)^{1/2}$ is the adiabatic sound speed, and $\Omega_k = (r\partial\Phi/\partial r)^{1/2}/r$ is the Keplerian angular velocity.

## 3 ROTATING STATIC TORI

### 3.1 Equilibria

We use the static tori as the initial condition of our time-dependent calculations. A static torus is described by equations (2.2)-(2.7) with time derivatives and all dissipative terms put to zero,

$$\frac{1}{\rho}\nabla P = -\nabla\Phi + \frac{\ell^2}{r^3}\nabla r. \quad (3.1)$$

According to the von Zeipel theorem, for a polytropic equation of state, $P = K\rho^\gamma$, where $K$ is the polytropic constant, the angular momentum depends only on the distance from the rotational axis, $\ell = \ell(r)$. In this case, equation (3.1) can be integrated as,

$$\frac{\gamma}{\gamma-1}\frac{P}{\rho} + W(r,z) = W_0 = \text{const}, \quad (3.2)$$

where $W(r,z)$ is the effective potential defined as

$$W(r,z) = \Phi(r,z) + \int_r^\infty \frac{\ell^2(r')}{r'^3}dr'. \quad (3.3)$$

The effective potential on the equatorial plane, $z = 0$, is shown in Figure 1a for the case of constant angular momentum, $\ell(r) = const$. The meridional cross section of equipotential surfaces are displayed in Figure 1b for $\ell = 2R_\text{G}c$. The boundary of a static torus has to coincide with one of the equipotential surfaces (Boyer 1965).

## 3.2 Stability

The Høiland criterion for static rotating fluid states that the fluid is stable against local, axisymmetric, adiabatic perturbations if and only if the following two conditions are satisfied (see Tassoul 1978 for details):

$$\frac{1}{r^3}\frac{\partial \ell^2}{\partial r} - \left(\frac{\partial T}{\partial P}\right)_S \nabla P \cdot \nabla S > 0, \quad (3.4a)$$

$$-\frac{1}{\rho}\frac{\partial P}{\partial z}\left(\frac{\partial \ell^2}{\partial r}\frac{\partial S}{\partial z} - \frac{\partial \ell^2}{\partial z}\frac{\partial S}{\partial r}\right) > 0. \quad (3.4b)$$

Here $S$ is the specific entropy. Note that flows in which the surfaces of constant specific entropy coincide with those of constant specific angular momentum are marginally stable with respect to the Høiland criterion (3.4b).

The static tori have been found by Papaloizou & Pringle (1984) to be unstable with respect to the *global non-axisymmetric*, dynamical modes. However, even a modest mass loss from the unstable tori suppresses the instability (e.g., Blaes 1987; Hawley 1987; Blaes & Hawley 1988; see also a review by Narayan & Goodman 1989).

Small scale *local non-axisymmetric* instabilities produce turbulence, for example, in the presence of a weak magnetic field (Balbus & Hawley 1991, Hawley & Balbus 1991). In this paper we use the $\alpha$-viscosity to model the small scale turbulence of an unspecified nature. It could be due to the instability discussed in the above quoted papers. The effective viscosity parameter induced by this instability is not well estimated, ranging from $\alpha \sim 0.1$ (Hawley, Gammie, & Balbus 1995) to $\alpha \sim 0.01$ (Brandenburg et al. 1995).

## 4 NUMERICAL TECHNIQUE

### 4.1 The Code

We use an explicit second-order Eulerian method to solve the time dependent equations for accretion flows. It is a modification of the Godunov (1959) scheme and is based on the hydrodynamical algorithm described in Harten et al. (1987) and Harten & Osher (1987). In this method, a representation of the hydrodynamical equations is made in which all the functions are approximated by piecewise linear distributions. The state of the system at the moment $t + \triangle t$ is determined in terms of the known state at moment $t$. The fluxes of density, momentum and energy on the boundaries of numerical cells are calculated at moment $t + \triangle t/2$ by solving the Riemann problem (van Leer 1979). No artificial viscosity is explicitly used.

We have taken into account all the viscous terms. In the regions of a large density gradient, the viscosity limited procedure is used to avoid strong unphysical torques. Specifically,

$$\mu' = \mu[\max(1, \triangle|\nabla \ln \rho|)]^{-2},$$

where $\triangle$ is the gridsize. Under this condition, the Reynolds number of the flow is always larger than unity in the regions of large density gradients. In the numerical scheme, the viscous terms are incorporated by the operator splitting method in which the first step deals with the hydrodynamical calculation and the second step deals with the viscous contribution. The second step is performed explicitly by using the state of the system calculated in the first step.

Extensive tests of the code have been made by comparing the numerical calculations with the analytic solutions of three well-known problems: decay of discontinuity, spherical Bondi accretion, and Sedov-type explosion.

### 4.2 Numerical Grid and Boundary Conditions

*The numerical grid* covers the calculation domain which is located inside $0 \leq r \leq r_B$ and $0 \leq z \leq z_B$, where $r = r_B$ and $z = z_B$ defines the outer boundary. We have used homogeneous grids in both radial and vertical directions. The gridsize is $\triangle r = \triangle z = 0.1 R_\mathrm{G}$, except in one case where a double resolution $\triangle r = \triangle z = 0.05 R_\mathrm{G}$ is used. The inner boundary around the black hole is defined at the sphere of radius $R_\mathrm{in}$ which corresponds to about 14 gridsize in the single resolution calculations. We assume axial symmetry with respect to the axis $r = 0$ and mirror symmetry with respect to the plane $z = 0$.

*The inner boundary condition* must be consistent with the hydrodynamics in the vicinity of the central black hole. We use a sucking inner boundary condition in which the fluid passes through the inner boundary freely and there is a negligible influence of the pressure from the inner layers. This kind of the sucking inner boundary condition was used in numerical calculations of accretion flows into a compact object by several investigators (e.g., Hunt 1971; Eggum et al. 1988; and Igumenshchev, Illarionov, & Kompaneets 1993).

The sucking boundary condition is adequate only if the boundary lies in the supersonic region, since in this case perturbations at the boundary cannot propagate outwards. In the case of a black hole accretion, the transition from subsonic to supersonic flow is expected to occur at $2R_{\rm G} \lesssim r_s \lesssim 3R_{\rm G}$ in the equatorial plane $z = 0$ (Abramowicz & Zurek 1981). Thus, by setting the sucking boundary condition at $R_{\rm in} \approx 1.3R_{\rm G} < r_s$ we should be able to adequately model the flow near the black hole.

*The outer boundary condition* should allow the possibility of both outflow and inflow of matter through the boundary. We have applied different outer boundary conditions for two different types of models. The first type of models has an initially small torus ($10R_{\rm G} - 15R_{\rm G}$) which is entirely contained in the calculation domain. In the second type of models, only the inner part of the torus is contained in the calculation domain. The first type of models is calculated to examine the global dynamical evolution of torus according to viscous redistribution of angular momentum. The second type of models is used to study the transonic nature of the flow. In both types of models, the free outflow condition without viscosity is set if the normal component of the fluid velocity in a boundary cell directs outward. In the opposite situation we use the inflow boundary condition, which is determined by the Bondi (1952) spherical accretion solution with the mass accretion rate as small as numerically possible. It should be stressed that the use of the Bondi solution is only a part of the numerical method and it does not influence the physics of the model.

In the second type of models, however, a different outer boundary condition is used if and when the boundary lies inside the torus. The surface of the torus is determined by $M$, $\ell_0$ and $W_0$ (see eq. [3.2]). For convenience, we use fictional cells which border the calculation domain. When they are located inside the torus, then at the beginning of each time-step, we reset the values of $v_r$ and $v_z$ to zero in these fictional cells, and restore the original values of $\rho$, $\ell$, $\varepsilon$ according to the static torus solution. Such boundary condition allows the plasma to either outflow or inflow through the boundary and it shall agree with the real flow if the velocity of plasma near the boundary is much less than the speed of sound. This was confirmed by several test calculations.

## 5 NUMERICAL RESULTS

In all models, an adiabatic index of $\gamma = 4/3$ is used, which agrees with radiation pressure dominated fluid. The mass of the central black hole is $M = 10\ M_\odot$. Rescaling results for different values of the mass $M$ (and the polytropic constant $K$) is trivial. In physical units, one has:

$$r_{\rm new} = \left(\frac{M}{10M_\odot}\right) r_{\rm old}, \quad t_{\rm new} = \left(\frac{M}{10M_\odot}\right) t_{\rm old}, \quad \ell_{\rm new} = \left(\frac{M}{10M_\odot}\right) \ell_{\rm old},$$

$$\rho_{\rm new} = \left(\frac{K_{\rm new}}{K_{\rm old}}\right)^{-\frac{1}{\gamma-1}} \rho_{\rm old}, \quad \dot{M}_{\rm new} = \left(\frac{M}{10M_\odot}\right)^2 \left(\frac{K_{\rm new}}{K_{\rm old}}\right)^{-\frac{1}{\gamma-1}} \dot{M}_{\rm old}.$$

In the rest of the paper, however, except where is indicated, we use $R_{\rm G}/c$, $c^2/2$, and $L_{Edd}/c^2$ as the units of time $t$, potential $W$, and mass accretion rate $\dot{M}$ respectively.

### 5.1 Models of Viscous Accretion Flows: Convective Instability

The initial non-accreting static torus is characterized by the constant angular momentum $\ell_0 = 2R_{\rm G}c$, the polytropic constant $K = 10^{20}$ in the $CGS$ units, and the surface potential $W_0 = -0.05$. Thus, initially, there is an effective potential barrier $\triangle W = -0.05$, and therefore accretion is not possible (see Figures 1ab). The calculation domain has the size of $0 \leq r \leq 20R_G$, $0 \leq z \leq 10R_G$, and the entire torus is contained in this domain. During the evolution of the models, no mass accretion is supplied at the outer boundary. Our goal here is to investigate the dynamics of the accretion flow after we 'switch on' the viscous redistribution of angular momentum.

The time-dependent behaviour of the models can be understood in the framework of the static torus approach (Section 3). The transport of angular momentum outward reduces the value of the angular momentum in the inner part of the torus, and consequently, the effective potential at the potential barrier, $W_*$, decreases. As a result, the local energy gap becomes $\triangle W = W - W_* > 0$ at some moment, and then accretion begins.

We describe here in detail numerical results for two particular models with $\alpha = 10^{-2}$ ($Model$ 1) and $\alpha = 10^{-3}$ ($Model$ 2).

$Model$ 1. Figure 2 shows the time evolution of the accretion rate $\dot{M}$ measured at the inner boundary from time $t = 0$ when the viscosity was 'switched on'. Before $t \simeq 70$ there is no significant accretion. After this moment accretion starts and continues with an increasing rate. The mass accretion rate reaches a maximum of $\dot{M}_{\max} \simeq 6.7$ at $t_1 \simeq 300$ and then it decreases slowly. The time evolution of the radial distribution of the specific angular momentum $\ell(r)$ at the equatorial plane is shown in Figure 3. Here, the corresponding Keplerian angular momentum $\ell_k = r(r\partial\Phi/\partial r)^{1/2}$ is displayed (the dashed line) for comparison. Note that the specific angular momentum exceeds the Keplerian one only in the innermost region of the disc, where the centrifugal force exceeds the gravitational one.

It is seen that after $t = t_1$ the distribution of specific angular momentum in the inner region of the accretion disc ($R \lesssim 7\, R_G$) does not change much (see curves 6, 7, 8, and 9 in Figure 3). The other parameters of the flow in this region also remain almost unchanged. Thus, the time $\tau_{st} = t_1$ characterizes the time scale of reaching the stationary state in the inner region of the disc. This time is proportional to the viscosity parameter $\alpha$, and depends on $\ell_0$ and $W_0$ in a more complicated way. The outer part of the initial torus slowly expands outward due to the redistribution of angular momentum. It outflows freely through the outer computational boundary. However, the rate of the mass loss through the outer boundary is much less than $\dot{M}$, and the influence of this mass loss on the flow is negligible.

The minimal accretion time in our model is $\tau_{accr} = M_{torus}/\dot{M}_{\max} \simeq 1.6 \cdot 10^3 \approx 5\tau_{st}$, where $M_{torus}$ is the initial mass of the torus. During $\tau_{st} < t \lesssim 5\tau_{st}$, an almost stationary accretion flow in the inner region of the disc is expected. This quasi-stationary flow is determined by the local viscous processes and also by the conditions in the middle region ($r \simeq 10 - 15\ R_{\rm G}$) which feeds the inner one.

Indeed, until $t \simeq 10^3$ our model shows a smooth flow and the time evolution is quasi-stationary. We now describe in some detail the characteristics of the model at this stage (see Figures 4a-d, $t = 995$; only the inner part of the flow is shown). The black region ($R \leq R_{\rm in}$, where $R_{\rm in}$ is the location of the inner boundary) covers the unresolved part of the accretion flow near the black hole.

Figure 4a shows the density contours and the position of the sonic surface (the heavy line). The contour lines are spaced with $\triangle \log \rho = 0.5$ and the largest value of $\log \rho$ is indicated in the units of $g\ cm^{-3}$. The sonic surface is defined as $\mathcal{M} = \sqrt{v_r^2 + v_z^2}/c_s = 1$, and the inflow is supersonic behind this surface (i.e., closer to the black hole). In agreement with the the analytic prediction (Abramowicz & Zurek 1981), the location of the sonic point $r_s$ at the equatorial plane is always slightly closer to the black hole than the left-hand intersection of curves $\ell(r)$ and $\ell_k(r)$ (see Figure 3).

Figure 4b shows the momentum vector field. It is convenient to show vectors of momentum $\rho\vec{v}$ rather than vectors of velocity $\vec{v}$ to illustrate the internal motion of matter. The scale for these vectors (the length of the arrow) is shown at the left-up corner of the figure. The inflow of matter takes place in a wide layer close to the surface of the disc and has a cone-shaped form. There is a stagnation point at $r \simeq 5.5\ R_{\rm G}$ in the equatorial plane. Close to the black hole the matter moves only inward. Away from the stagnation point there is an outflow which takes place near the equatorial plane. A similar flow pattern was previously found by Urpin (1984) and Siemiginowska (1988) for thin accretion discs. Their results were later confirmed by Kley & Lin (1992) and Różyczka et al. (1994) in two-dimensional models.

The contours of specific angular momentum are shown in Figure 4c. The heavy line represents the value of the initial constant specific angular momentum. Note that $\ell(r,z)$ increases with $r$, but depends on $z$ weakly at distances larger than $r \simeq 5\ R_{\rm G}$. The redistribution of angular momentum is due to the viscosity.

The contours of specific entropy are presented in Figure 4d. Initially, the specific entropy is constant inside the torus. The increase of entropy is due to the dissipative viscous processes. Near the surface of the disc the viscous processes are less significant and the specific entropy has a local minimum there. Inside the disc, the specific entropy increases towards the equatorial plane and has a local maximum near the sonic point $r_s$. In the supersonic region of the flow, however, the viscous processes are not important, and consequently, the specific entropy is almost constant.

Note that the model at this stage remains stable even though it is easy to check from Figures 4cd that the Høiland criterion (3.4b) is not satisfied (but condition (3.4a) is satisfied). This contradiction can be explained by the stabilization provided by the large scale motion of matter.

The subsequent evolution of the model shows that the accretion flow becomes convectively unstable, and the flow pattern is changing significantly at distances $r \gtrsim 6\ R_{\rm G}$. One can observe the initial stage of the development of the convective instability in Figures 5ab, where the contours of specific angular momentum and specific entropy, and also the momentum vector field, are presented ($t = 1481$). The emergence of a hot bubble (which is the region of locally increased specific entropy) is clearly seen at distance $\simeq 7 - 8\ R_{\rm G}$ in the equatorial plane. The bubble is accompanied by two ring eddies.

The matter of the bubble is less dense than that of the surrounding gas. Under the action of the Archimedes buoyancy force, the toroidal convective bubble flows outwards along the equatorial plane. Shortly later ($t = 1849$), the bubble has visibly expanded, and the size of the accompanying eddies became larger as well (Figures 6ab).

At this moment, the Høiland criterion (3.4b) is not satisfied everywhere. Nevertheless, this fact does not imply global instability when a significant large scale motion of matter takes place. In Figures 5ab and 6ab, regions unstable with respect to the Høiland criterion (3.4a) are displayed as the grey areas. They are only partially associated with the convectively unstable regions. For example, although there is always a grey area at distances $r \lesssim 5\ R_G$ no global convective instability occurs, due to the powerful stabilizing action of accretion which takes place there. In the other parts of the flow, the Høiland criterion however gives correctly the location of the unstable region which is always connected with the presence of the convective hot bubbles.

Further calculations show the development of a complicated flow pattern. The perturbation due to the moving of the convective bubble propagates inwards. The development of the perturbations makes the inner region of the disc unstable by producing vortices (see Figure 7a, $t = 2493$). These vortices mix the matter efficiently, and redistribute angular momentum and entropy. In Figure 7a the regions unstable according to the Høiland criterion (3.4ab) are shown as the grey areas. It is interesting to see that the mixing process also produces regions which are stable against local axisymmetric perturbations (nonshadow areas). Later (Figure 7b, $t = 2890$), these stable regions predominate in the disc and the vortex activity is less efficient.

The vortex activity inside the disc affects the accretion rate $\dot{M}$. This can be seen in Figure 2, which shows that non-monotonic behaviour of $\dot{M}(t)$ takes place during the time of $t \approx 2400 - 2800$ when maximum vortex activity in the disc is observed.

The midplane outflow is always present as a background flow, and the convective bubble discussed above moves in this background. The bubble originates near the stagnation point and provides a substantial perturbation in the background outflow. Due to the computer time limitation we could not follow the further evolution of the disc. Most probably, the pattern of this behaviour will repeat and the transient convective bubbles will quasi-periodically emerge from the stagnation point. However, the time scale to form another bubble must be longer than the period of the accretion rate oscillations observed in *Model* 2 (see below).

*Model* 2 ($\alpha = 10^{-3}$). This model differs both qualitatively and quantitatively from *Model* 1. The main qualitative difference is that in the *Model* 2 the flow is never smooth. Meridional circulations and vortices are present in the inner part of the torus right from the start of accretion. Time-dependent meridional circulations produce quasi-periodic oscillations in $\dot{M}$ with a time scale of $\sim 900$ (or $\sim 0.01\ M/M_\odot$ in seconds) and with an amplitude of $\sim 20$ % (see Figure 8). We have performed a FFT power spectrum analysis for $\dot{M}(t)$ between $t = 1000$ and $4300$, and observed a broad peak at frequency about $10^{-3}$ Hz and also some harmonics. It should be reminded however that the entire evolution time is rather short considering the QPO time scale involved and thus caution should be used to draw any quantative conclusions.

In Figures 9abcd we present the flow patterns at four different moments corresponding to the minima and maxima on the curve $\dot{M}(t)$ which are indicated in Figure 8 by arrows and letters. It is clear that the flows with the more efficient mixing of matter are associated with the larger accretion rate and vice versa. From this behaviour it should be obvious that models with small viscosity are unstable. In Figures 9abcd the regions which are unstable according to the Høiland criterion (3.4ab) are shown as the grey areas. Contrary to what we have seen in *Model* 1, the large scale motion of matter does not stabilize the instability in the bulk of the disc, accept in the region $r \lesssim 4\ R_{\rm G}$. In the nonlinear regime of the instability, the accretion flow produces a significant meridional circulation and vortices with various length-scales: from the maximal one which is comparable to the local thickness $H$ of the disc, to the minimal one which is comparable to the numerical resolution.

The instability is saturated when the condition (3.4b) is marginally fulfilled. This happens at later times when the surfaces of constant specific angular momentum $\ell = const$ and entropy $S = const$ almost coincide.

Properties of the model at the last calculated moment $t = 4335$ are presented in Figures 10abcd. The density contours and the position of the sonic surface are shown in Figure 10a. The flow pattern and the location of the unstable (grey) regions are displayed in Figure 10b. Considerable part of the disc is stable according to the Høiland criterion (3.4ab). In the region $r < 6\ R_G$ there is a very efficient meridional circulation. In other parts of the disc, vortices with different length scales are present, but the efficiency of the mixing of matter is low at this stage. The distribution of angular momentum $\ell$ and specific entropy $S$ are shown in Figures 10cd. One sees that the surfaces of constant specific angular momentum and entropy almost coincide, which is consistent with the marginally stable state of the flow.

The fact that in the flows which are convectively unstable the surfaces of constant specific angular momentum and entropy must coincide ($\nabla l \times \nabla S = 0$, implies marginal stability according to Høiland criterion [3.4b]) was anticipated by Bardeen (1973) and Paczyński & Abramowicz (1982).

We note that our viscous accretion disc models are characterized by the weak fluxes of the 'evaporated' matter from the surface of the disc. This process is due to viscosity. Evaporated matter has a positive value of the Bernoulli parameter, which is the sum of the kinetic and gravitational energies per unit mass, and the specific enthalpy of the gas. This is why the matter moves outward with almost virial velocities along the surface of the disc, and forms a dynamical disc corona. This mechanism of the origin of outflow from accretion discs was pointed out by Narayan & Yi (1994, 1995).

## 5.2 Models of Isentropic Accretion Flows

In the innermost regions of the disc the dissipation of angular momentum plays no important role and consequently in our models the viscosity is put to zero. Accretion occurs due to a slight overflow of the effective potential barrier at $r = r_*$, $z = 0$ as discussed in Kozłowski et al. (1978).

The calculation domain for this type of models is $0 \leq r \leq 10R_G$ and $0 \leq z \leq 10R_G$. We start the calculation from the initially non-accreting torus which contains the calculation domain (see Section 3). The polytropic constant is $K = P/\rho^\gamma = 10^{22}$ in $CGS$ units, and $\ell_0 = 2R_G c$. For this value of $\ell_0$ the maximum effective potential barrier is $W_* = W(r_*, 0) = 0$ and one needs to set $W_0 > 0$ to allow accretion. With this choice, matter close to the center of the black hole ($r < r_* = 2R_G$) cannot be in a stable equilibrium. We have calculated models with $W_0 = 0.01, 0.02, 0.04, 0.08$, and $0.16$. The calculation of each model was continued until a steady state was reached. Typically, at that time, all the mass initially contained in the calculation domain was accreted into the black hole.

Each numerical model is characterized by the final stationary accretion rate $\dot{M}$ (in physical unit). They are shown as a function of the energy gap $\triangle W = W_0 - W_*$ in Figure 11. The full circles represent single resolution models and the open circle is for the double resolution model. The solid line represents the analytic solution of Abramowicz (1985) which can be written as

$$\dot{M} = A(n) K^{-n} \frac{r_*}{\Omega_k(r_*, 0)} (\triangle W)^{n+1}, \qquad A(n) = \frac{(2\pi)^{3/2} n^n \Gamma(n + 3/2)}{(n+1)^n (n+1/2)^{n+1/2} \Gamma(n+2)}. \qquad (5.1)$$

Here $\Gamma$ is the Euler gamma function and $n = 1/(\gamma - 1)$ is the polytropic index. The agreement between analytic and numerical results is very good. This shows that formula (5.1) may be used as a convenient accuracy check of numerical calculations in the pseudo-Newtonian potential. The model with $W_0 = 0.01$ was recalculated with double spatial resolution.

We show now in some more details the behaviour of two models characterized by $W_0 = 0.01$ with double resolution and $W_0 = 0.16$. The density contours and momentum vector field are shown in Figures 12ab and 13ab respectively. As in *Model* 1, the density on the contour lines increases toward the mid-plane and the lines are spaced with $\triangle \log \rho = 0.5$. The model which has the larger energy gap ($\triangle W = 0.16$) is vertically thicker ($\simeq 1\ R_{\rm G}$) in the transonic region, and its sonic point is located at $r_s = 1.9\ R_{\rm G}$ in equatorial plane (see Figure 13ab). The model with $\triangle W = 0.01$ is vertically thinner and has $r_s = 2.0\ R_{\rm G}$ (see Figure 12ab).

From Figures 12b and 13b, it is seen that the fluid motion inside the torus corresponding to the model with a small energy gap is complicated and resemble a circulation: many stream lines originate and terminate on the outer boundary of the calculation domain and only a small part of matter goes into the black hole. The circulation is essentially subsonic. In the case of a large energy gap, most of the matter moves along the stream lines which originate on the outer boundary and goes directly to the inner boundary. The difference of the flow patterns in these two models can be explained as the following. The models are marginally stable according to the Høiland criterion (3.4ab). The small numerical diffusion which is intrinsic for any Eulerian hydrodynamical code may slightly destabilize the marginally stable state. As a result, a weak circulation of matter takes place for the small energy gap model where no strong radial motion is present. The other model has a strong radial inflow, which stabilizes this instability.

To examine the influence of the numerical diffusion (or numerical viscosity) on our results, we estimate the numerical diffusion by its effective Reynolds number (Oran & Boris 1987) as,

$$\Re_{num} \approx \frac{2L}{c_1 \triangle},$$

where $L$ is the length-scale of the flow and $\triangle$ is the gridsize. The value of parameter $c_1$ depends on the type of the numerical scheme. In the case of nonlinear monotonic schemes, such as the one used here, $c_1 \simeq 10^{-1}$ if the flow is smooth. Thus, in the interior of the torus we have $\Re_{num} \simeq 2000$ which corresponds to an effective viscosity parameter,

$$\alpha_{num} = \frac{1}{\Re_{num}} \left(\frac{\triangle}{H}\right)^2 \sim 10^{-7}.$$

Near the surface of the disc the gradients of all the variables are very large. For example, the relative jump of the density there is $\gtrsim 10^8$. Our scheme is able to avoid the numerical diffusion even in this kind of extreme cases: we can keep the contact discontinuity and also satisfy a consistent advection of the hydrodynamical variables. Test calculations show that only a negligible flux of the 'evaporated' matter from the surface of the disc was present during the long time calculations.

## 6 DISCUSSION AND CONCLUSION

We have constructed numerically several models of transonic accretion flows near a black hole. We assumed that the radiative losses of heat are negligible, and that the flow is mostly cooled by advection. Our models are two dimensional (axially symmetric) and time-dependent. We model the gravitational field of the black hole by the pseudo-Newtonian potential. We assume that (unspecified) instabilities generate a small-scale turbulence which we describe by the $\alpha$-viscosity prescription. We take into account all the viscous effects of heat generation and redistribution of angular momentum. We follow the evolution of the accretion flow from the initially non-accreting state which is described by a static thick torus rotating around the black hole. We found that the flow pattern is rather complex due to the developement of a convective instability. This is due to the fact that advection dominated accretion flows are convectively unstable as pointed out previously by Narayan & Yi (1994, 1995). The behaviour of flows with a small assumed viscosity differs from the behaviour of flows with large viscosity.

In the case of a small viscosity, the instability is not suppressed by the accretion flow, as the accretion flow is not very strong. Thus, the flow pattern is very noisy all the time. Convective bubbles and ring-like eddies are constantly formed. The model with the small value of $\alpha = 10^{-3}$ shows development of a complicated flow pattern with no phase of smooth accretion flow. The resulted distributions of specific entropy and angular momentum are unstable according to the Høiland criterion (3.4b), and small perturbations develope in the non-linear regime into multi-vortex structures and meridional circulations. These motions redistribute the entropy and angular momentum in a way which helps to remove the reason of instability. According to the Høiland criterion (3.4b) the marginally stable state is reached when the surfaces of constant specific entropy and angular momentum coincide. In our models, the resulting distribution eventually locally approaches this marginally stable state. This property of accretion discs was predicted by Paczyński & Abramowicz (1982). It should be noted that in this particular case the convective motion transfers angular momentum outwards to produce an effective viscosity parameter larger than the original one.

In the case of a high viscosity, the non-linear effects of strong accretion flow smooth the convective instability. The model with the large value of $\alpha = 10^{-2}$ shows, after a short switch-on period, a smooth, quasi-stationary accretion. The flow is globally stable even though the Høiland criterion (3.4b) is not satisfied. It consists of the inflow in a wide layer close to the surface of disc and the outflow near the equatorial plane starting at the distance $\simeq 5.5\ R_\mathrm{G}$ from the center. Further evolution, however, brings in a convective instability. We observed the formation of a big, hot bubble as a perturbation in the equatorial outflow. The bubble moves outwards, accompanied by big, ring-like eddies. Later, a more complicated flow pattern emerges. The secondary perturbation caused by the moving bubble propagates to the inflow region and makes it unstable. As a result, the whole inner region of the disc is filled with vortices of different length-scales. The vortex structure observed in our calculations is two-dimensional, and correspondently, we observe the process of confluence of smaller eddies forming larger eddies. Obviously, in reality the vortex structure must be three-dimensional for such kind of flows. Most probably, there will be a 3D-turbulisation of the viscous accretion flow due to the convective instability when a cascade of energy from the larger scales to the smaller scales takes place. This turbulence may explain the nature of the viscosity in the advection dominated accretion discs.

Our viscous accretion disc models are not fully satisfactory as models of stationary disc accretion, because we study the viscous evolution of an isolated torus with a limited mass. However, it is unlikely that this may influence the general properties of hydrodynamical instabilities found in the present work. Due to the angular momentum redistribution a part of the torus moves outwards, but the main part is accreted into the black hole. The characteristic time of this process is the accretion time of the torus. It is much longer than the hydrodynamic instabilities which develope in the accretion flow. Thus, one may assume that these instabilities will also be present in a more realistic case, when mass is constantly being added to the torus.

The temperature in the inner part of our disc models is very high, $T \sim 10^{11} K$. Thus, the microphysical effects in the relativistic plasma (e.g., electron-positron pairs) become important. Our simplified models do not describe these processes. We approximate the relativistic plasma in the classical fluid dynamics assuming $\gamma = 4/3$. Furthermore, it is more likely that the radiative cooling may introduce some changes, and in a future study we shall investigate how the picture presented here may be affected when one includes the radiative cooling. An interesting problem is the influence of the rotation of the black hole on the transonic region of the accretion disc. But, this investigation must use the relativistic hydrodynamics approach in the Kerr geometry.

It is worthwhile to mention that the unstable nature of accretion flow near the black hole may provide an explanation to the quasi-periodic oscillation phenomena (QPO) observed in the Galactic black hole candidate X-ray sources. In the case of *Model* 2, the instability introduces a low amplitude (20%) oscillations which have a typical frequency about 100 $M_\odot/M$ Hz. This is comparable to the high frequency ($\sim 4 - 10$ Hz) QPOs of black hole candidates if we assume the mass of the black hole is about 10 $M_\odot$. Preliminary results also suggest that the the QPO frequency is approximately proportional, in a range of, to the viscosity parameter $\alpha$. Detailed parameter study will be our future work.

The authors are grateful to Axel Brandenburg, Andrei Illarionov and Ed Spiegel for discussions.

28## FIGURE CAPTIONS

**Figure 1.**— (a) The distribution of the effective potential on the equatorial plane $W(r,0)$ for different constant values of angular momentum, $\ell(r) = const$. Here $\ell_{mb} = 2\,R_\mathrm{G} c$ and $\ell_{ms} \approx 0.9186\,\ell_{mb}$ are the specific angular momenta of marginally bound and marginally stable orbits respectively. b) The meridional cross section of the equipotential surfaces $W(r,z) = const$, for the case of $\ell = \ell_{mb}$. The equipotential lines are spaced with $\triangle W = 0.025$ in units of $c^2/2$.

**Figure 2.**— The evolution of accretion rate $\dot{M}$ for $Model$ 1 with viscosity parameter $\alpha = 10^{-2}$. The time $t$ and $\dot{M}$ are given in the units of $R_G/c$ and $L_{Edd}/c^2$ respectively.

**Figure 3.**— The evolution of the distribution of specific angular momentum $\ell$ at the equatorial plane of $Model$ 1. The solid lines labeled from 1 to 9 correspond to the moments $t = 0$, 32.1, 59.5, 83.4, 133.8, 260.2, 506.1, 994.7, and 1357.0 respectively. The Keplerian angular momentum $\ell_k$ is displayed as the dashed line for comparison.

**Figure 4.**— The flow pattern of $Model$ 1 at time $t = 994.7$. The black region ($R \leq R_\mathrm{in}$, where $R_\mathrm{in}$ is the location of the inner boundary) covers the unresolved part of the accretion flow near the black hole. a) The contours of density. The contour lines are spaced with $\triangle \log \rho = 0.5$ and the maximum $\log \rho$ is indicated in the units of $g\ cm^{-3}$. The sonic surface (Mach number $\mathcal{M} = 1$) is represented by the heavy line. b) The momentum vector field. The direction and the length of the arrow show the momentum vector $\rho \vec{v}$. The scale for these vectors is shown at the left up corner of the figure. c) The contours of specific angular momentum. The heavy line corresponds to the initial value of the angular momentum, $\ell_0 = 2R_\mathrm{G} c$. The lines are spaced with $\triangle \ell = 0.01\,\ell_0$. d) The contours of specific entropy. The local maximum of the specific entropy is located near $r \approx 2.5\,R_\mathrm{G}$, $z = 0$.

**Figure 5.**— The flow pattern of $Model$ 1 at time $t = 1481.2$. A convective hot bubble is seen to emerge. The grey areas are the regions unstable according to the Høiland criterion

(3.4a). a) The contours of specific angular momentum (solid lines) and entropy (dashed lines). b) The momentum vector field.

**Figure 6.**— Same as Figure 5 but at time $t = 1849.2$. The convective bubble has visibly expanded.

**Figure 7.**— The momentum vector field. The regions which are unstable according to Høiland criterion (3.4ab) are indicated as grey areas. (a) $t = 2493$. The flow pattern becomes complicated due to the perturbation of the hot bubble which is moving outward. The flow is unstable. (b) $t = 2890$. The flow pattern is still complicated but stable regions (nonshadow areas) predominate in the disc and the vortex activity becomes less efficient.

**Figure 8.**— The evolution of accretion rate $\dot{M}$ for $Model$ 2 with viscosity parameter $\alpha = 10^{-3}$. The disc is unstable and quasi-periodic oscillations are clearly seen. The labels indicate the evolution stages for which detailed flow patterns are presented in Figure 9.

**Figure 9.**— The momentum vector field. The regions which are unstable according to Høiland criterion (3.4ab) are indicated as grey areas. (a)-(d) correspond to four different times which are labeled in Figure 8. Models with the more efficient mixing of matter are associated with the larger accretion rates. The efficient mixing of matter produces stable regions in the disc.

**Figure 10.**— Same as Figure 4 but for $Model$ 2 at time $t = 4334.9$. Grey areas in (b) are the unstable regions according to the Høiland criterion (3.4ab).

**Figure 11.**— The mass accretion rate $\dot{M}$ for isentropic accretion as a function of the energy gap $\triangle W$. The full circles represent single resolution models and the open circle is for the double resolution model. The analytic solution (5.1) is shown by the solid line.

**Figure 12.**— The stationary flow patterns for the isentropic flow. Here the energy gap is $\triangle W = 0.01$ in units of $c^2/2$. (a) The contours of density. (b) The momentum vector field. The sonic surface (Mach number $\mathcal{M} = 1$) is represented by the heavy line.

Figure 13.— As in Figure 12, but for $\triangle W = 0.16$.